\documentclass[
    ,final            
  ,numberedheadings 
  ]
  {aipproc}

\usepackage[mathscr]{eucal}
 \usepackage{amsmath, amssymb}

\def\dmf{\dot{\mathfrak{M}}}
 \def\msE{\mathscr{E}}
  \def\sun{\hbox{$\odot$}}
   
\newcommand{\be}{\begin{equation}}
\newcommand{\ee}{\end{equation}}
\newcommand{\bdm}{\begin{displaymath}}
\newcommand{\edm}{\end{displaymath}}
\layoutstyle{6x9}

\begin{document}

\noindent {\it AIP Conference Proceedings, Vol.~1439, pp.~237-248 (2012)}\\
Proc. of WISAP\,2011 Conference ``Waves and Instabilities in Space and Astrophysical Plasmas'',
P.-L.\,Sulem \& M.\,Mond (eds.), Eilat, Israel,  June 19th  -  June 24th, 2011

\vspace{1cm}

\title{A new look at spherical accretion in High Mass X-ray Binaries}


\author{N.R.\,Ikhsanov}{
  address={Pulkovo Observatory, Pulkovskoe Shosse 65, Saint-Petersburg 196140, Russia}
}

\author{L.A\,Pustil'nik}{
  address={Israel Space Weather and Cosmic Ray Center, Tel Aviv University, Israel Space Agency, \& Golan Research Institute, Israel}
}

\author{N.G.\,Beskrovnaya}{
  address={Pulkovo Observatory, Pulkovskoe Shosse 65, Saint-Petersburg 196140, Russia}
}

\begin{abstract}
Currently used model of spherical accretion onto a magnetized rotating neutron star encounters major difficulties in explaining the entry rate of accreting material into the stellar field and spin evolution of long-period X-ray pulsars. These difficulties can be, however, avoided if the magnetic field of the material captured by the neutron star is incorporated into the model. The magnetic field of the flow itself under certain conditions controls the accretion process and significantly affects the parameters of the accreting material. The mode by which the accretion flow enters the stellar magnetosphere in that case can be associated with Bohm (or turbulent) diffusion and the torque applied to the neutron star appears to be substantially higher than that  evaluated in the non-magnetized accretion scenario.
\end{abstract}

\maketitle


 \section{Introduction}

High Mass X-ray Binaries (HMXBs) are interacting binary systems made of a massive O/B-type star and a magnetized, rotating neutron star. The massive star underfills its Roche lobe and looses material in a form of stellar wind. As the neutron star moves through the wind of density $\rho_{\infty}$ with a relative velocity $\vec{V}_{\rm rel} = \vec{V}_{\rm ns} + \vec{V}_{\rm w}$, it interacts in a unit time with the mass $\dmf_{\rm c} = \pi R_{\rm G}^2 \rho_{\infty} V_{\rm rel}$. Here $R_{\rm ns}$ and $M_{\rm ns}$ are the radius and the mass of the neutron star, $V_{\rm ns}$ is the velocity of its orbital motion, and $V_{\rm w}$ the wind velocity. The Bondi radius, $R_{\rm G} = 2GM_{\rm ns}/V_{\rm rel}^2$, represents the maximum distance at which the neutron star is able to capture material from its environment.

The captured material initially follows ballistic trajectories and approaches the neutron star to the distance $R_{\rm A} = \left(\mu^2/\dmf_{\rm c} (2GM_{\rm ns})^{1/2}\right)^{1/2}$ at which the magnetic pressure due to the dipole field of the star reaches the ram pressure of the accreting material. $R_{\rm A}$ represents the Alfv\'en radius and $\mu$ is the dipole magnetic moment of the neutron star. The accretion process inside the Alfv\'en radius is fully controlled by the stellar field. The accreting material penetrates into the field at the magnetospheric boundary and reaches the stellar surface flowing along the magnetic field lines to the magnetic pole regions. The accretion power is then converted into X-rays. If the star is obliquely rotating it appears as an accretion-powered pulsar. The period of pulsations in this case is equal to the spin period of the neutron star. The mass capture rate by the neutron star in a stationary accretion picture can be evaluated through observations of the source X-ray luminosity, $L_{\rm x}$, as $\dmf_{\rm a} = L_{\rm x} R_{\rm ns}/GM_{\rm ns}$.

The accretion flow beyond the magnetospheric boundary can be treated in spherically symmetrical or disk approximation. The specific angular momentum of the captured material is $k_{\rm a0} \sim \Omega_{\rm orb} R_{\rm G}^2$, where $\Omega_{\rm orb} = 2 \pi/P_{\rm orb}$ is the orbital angular velocity and $P_{\rm orb}$ the orbital period of the binary system. The transfer rate of angular momentum associated with accretion is $\dot{J} = \xi \Omega_{\rm orb} R_{\rm G}^2 \dmf_{\rm a}$, where $\xi$ is the factor by which angular momentum accretion rate is reduced due to inhomogeneities (velocity and density gradients) and magnetic viscosity in the accretion flow. The angular velocity of the material increases during accretion process from the initial value $\omega_0 = \Omega_{\rm orb}$ to $\omega_{\rm en}(r) = \xi \omega_0 \left(R_{\rm G}/r\right)^2$. As it reaches the Keplerian angular velocity, $\omega_{\rm k} = \left(GM_{\rm ns}/r^3\right)^{1/2}$, the accretion disk forms. The distance at which formation of the disk in a wind-fed binary system can be expected is referred to as the circularization radius, $R_{\rm circ} = \dot{J}^2/GM_{\rm ns} \dmf_{\rm a}^2$. It is defined by equating $\omega_{\rm en}(R_{\rm circ}) = \omega_{\rm k}(R_{\rm circ})$. The Keplerian disk around a magnetized neutron star could be formed only if $R_{\rm circ} > R_{\rm A}$. Otherwise, the stellar magnetic field would prevent the material from reaching the circularization radius. This inequality can be expressed as (see \cite{Ikhsanov-2007}, and references therein)
 \bdm
V_{\rm rel} < V_{\rm cr} \simeq 200\ {\rm km\,s^{-1}}\ \xi_{0.2}^{1/4}\ \mu_{30}^{-1/14}\ m^{11/28}\ P_{40}^{-1/4}\ \dmf_{17}^{1/28},
 \edm
where $\mu_{30}=\mu/10^{30}\,{\rm G\,cm^3}$ and $m$ is the mass of the neutron star in units of $1.4\,M_{\sun}$. The system orbital period, $P_{40}$, is expressed in 40\,days, and $\dmf_{17}= \dmf_{\rm c}/10^{17}\,{\rm g\,s^{-1}}$. Finally, $\xi_{0.2}=\xi/0.2$ is normalized to its maximum average value in the case of non-magnetized accretion flow (see e.g. \cite{Ruffert-1999}, and references therein).

\begin{table}
   \begin{tabular}{lccccccc}
        \noalign{\smallskip}
    \hline
            \noalign{\smallskip}
 Name & Sp. type\,$^\dag$ & $P_{\rm orb}$, d & $P_{\rm s}$, s & $\dot{P}$, s/s\,$^\ddag$ & CRSF & $\log{L_{\rm x}}$\,$^\sharp$ & d \\
        \noalign{\smallskip}
     \hline
             \noalign{\smallskip}
GX\,301--2 &  B1~Ia & 41.5 & 683 & $[\,\pm \,]~{\rm E}-7.3$ & $30-38$\,keV & 37--37.5 & 1.8--3\,kpc \\
     & & & & & & & \\
X~Persei  & B0~Ve  & 250 & 837 & $[\,\pm \,]~{\rm E}-5.5$ & $29$\,keV & 34.7--35.5  & 950\,pc \\
         & & & & & & & \\
4U\,2206+54 & O9e & 19.25 & 5554 & $[\,+\,]~{\rm E}-6.3$ &   & 35--35.6  & 2.6\,kpc \\
         & & & & & & & \\
2S\,0114+65 & B1~Ia & 11.6 & 10008 & $[\,-\,]~{\rm E}-6.1$ & & 35--36 & 3\,kpc \\
        \noalign{\smallskip}
    \hline
            \noalign{\smallskip}
 \multicolumn{8}{l}{$^*$~For references see \cite{Ikhsanov-2007}} \\
  \multicolumn{8}{l}{$\dag$~Spectral type of the massive star,} \\
    \multicolumn{8}{l}{$\ddag$~Spin behavior ($[\,-\,]$ for spin-up and $[\,+\,]$ for spin-down) and maximum absolute value of $\dot{P}$,} \\
  \multicolumn{8}{l}{$\sharp$~$L_{\rm x}$ is in erg\,s$^{-1}$} \\
   \end{tabular}
   \caption{Persistent Long-period X-ray Pulsars$^*$}
\label{t1}
 \end{table}

The wind velocity of O/B-type stars usually ranges between $400$ to $1000\,{\rm km\,s^{-1}}$. A smaller wind velocity can be expected only in Be-type stars, which are surrounded by a viscous outflowing disk. But even in that case the orbital velocity of neutron stars would exceed $V_{\rm cr}$ if the orbital period of the binary system satisfies the condition $P_{\rm orb} < 50$\,days. It, therefore, appears that a formation of Keplerian accretion disk in a large number of HMXBs is not expected and the accretion process in those systems can be considered in a quasi-spherical approximation.

\begin{figure}
  \includegraphics[height=.4\textheight]{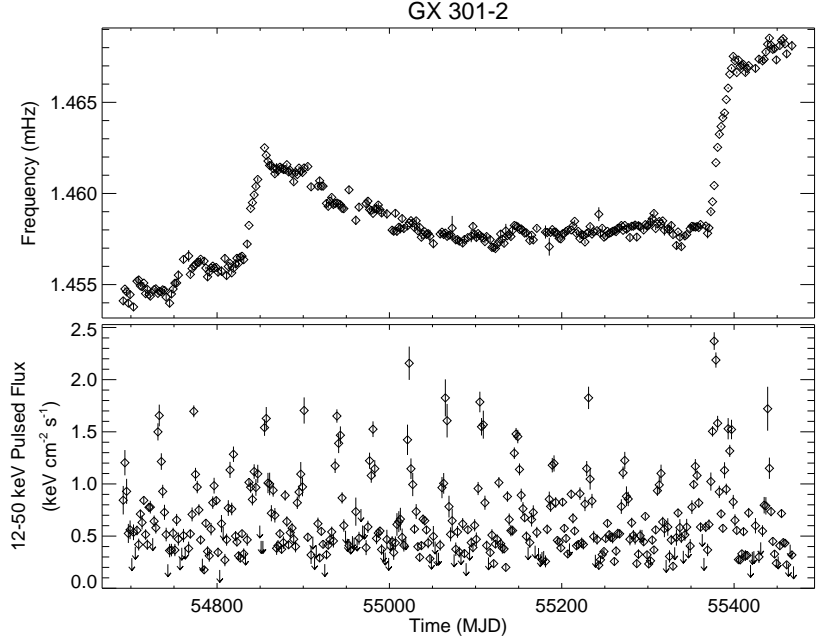}
  \caption{Pulse frequency and 12--50\,keV pulsed flux of GX\,301--2 observed with the Fermi Gamma-ray Burst
Monitor detectors. GMB Pulsar Project http://gammaray.nsstc.nasa.gov/gbm/science/pulsars}
   \label{f1}
\end{figure}

About 60 X-ray pulsars in our Galaxy are associated with HMXBs \cite{Liu-etal-2006}. More than 10 of them represent the subclass of persistent long-period X-ray pulsars. Parameters of the best studied members of this subclass are summarized in Table\,\ref{t1}. They are the wind-fed accretors. Observations give no evidence for a presence of persistent accretion disk in these systems. Instead, there are some indications that the neutron star in 2S\,0114+65 is embedded in and accreting from a dense low angular momentum gas cloud \cite{Masetti-etal-2006}. GX\,301--2 is the bright pulsar with the X-ray luminosity $(1-3) \times 10^{37}\,{\rm erg\,s^{-1}}$. The rest of the sources are fainter: $L_{\rm x} < 10^{36}\,{\rm erg\,s^{-1}}$. Nevertheless, the X-ray spectra of all these pulsars look similar, which strongly suggests that the accretion process operating in these binary systems is governed by the same mechanism. Observations of the cyclotron resonance scattering feature (CRSF) in X-ray spectra of GX\,301--2 and X~Persei suggest that the surface field of the neutron stars lies in the range of $(2-4) \times 10^{12}$\,G. All of these pulsars display a spin evolution. 2S\,0114+65 shows spin-up and 4U\,2206+54 spin-down behavior. Both the spin-up and spin-down phases have been observed in GX\,301--2 and X~Persei. The recent spin history of GX\,301--2 is shown in Fig\,\ref{f1}. The long-term spin-down trends at the rate $\dot{\nu}_{\rm sd} \simeq - 10^{-13}\,{\rm Hz\,s^{-1}}$ are superposed on slow spin-up trends and rapid spin-up episodes at the rate $\dot{\nu}_{\rm su} \simeq 5 \times 10^{-12}\,{\rm Hz\,s^{-1}}$ \cite{Lipunov-1982}, \cite{Koh-etal-1997}.

The accretion picture in the pulsars described above is the subject of our paper. We show that the currently used non-magnetized accretion flow scenarios encounter major difficulties in explaining the spin evolution and persistent character of the pulsars (next Section). Then we discuss the strength of the magnetic field in the accretion flow, elaborate the Magnetically Controlled Accretion (MCA) scenario and consider its application to the interpretation of GX\,301--2. A brief summary of our results is concluding the paper.

  \section{Non-magnetized accretion flow scenario}\label{non-ms}

The magnetic field of the accretion flow is usually neglected in currently used spherical accretion scenarios. It is assumed that the neutron star captures material at the rate $\dmf_{\rm a} = \dmf_{\rm c}$, which is then falling towards the star in spherically symmetric fashion with the free-fall velocity $V_{\rm ff}(r) = \left(2GM_{\rm ns}/r\right)^{1/2}$. The gravitational energy of the star is converted predominantly into kinetic energy of the accreting material and its ram pressure increases as
 \be\label{eram}
\msE_{\rm r}(r) = \msE_{\rm r0} \left(\frac{r}{R_{\rm G}}\right)^{-5/2},
 \ee
where $\msE_{\rm r0} = \rho_{\infty} V_{\rm rel}^2$ is the ram pressure at the Bondi radius.

The flow is decelerated by the stellar magnetic field when it approaches the star to a distance of $\sim R_{\rm A}$. This leads to the formation of a magnetosphere of radius $R_{\rm m}$ and to the heating of the flow up to a temperature $T_{\rm s} = (3/16) T_{\rm ff}(R_{\rm m})$, where $T_{\rm ff}(r) = GM_{\rm ns} m_{\rm p}/k_{\rm B} r$ is the proton free-fall temperature, $m_{\rm p}$ is the proton mass and $k_{\rm B}$ is the Boltzmann constant \cite{Lamb-etal-1977}. The heating occurs on the dynamical (free-fall) time, $t_{\rm ff}(r) = \left(r^3/2 GM_{\rm ns}\right)^{1/2}$, in the shock located at the magnetospheric boundary. The magnetosphere surrounded by the hot spherically symmetric accretion flow is closed (i.e. does not contain open field lines). It is rotationally symmetric about the magnetic axis and reflection symmetric about the equatorial plane. The boundary of the magnetosphere is convex towards the accreting material and contains two cusp points situated on the stellar magnetic axis. The size of the magnetosphere in the plane of the magnetic equator is $R_{\rm m} \simeq (1.7 - 2) \times R_{\rm A}$ and the distance to the cusp points is close to the Alfv\'en radius of the neutron star (see \cite{Arons-Lea-1976a}, \cite{Elsner-Lamb-1977}, \cite{Michel-1977}).

Formation of the magnetosphere, in the first approximation, prevents the accreting material from reaching the stellar surface. To enter the magnetosphere the material should move across the magnetic field lines. However, the rate of diffusion, $\dot{M}_{\rm in} \sim 2 \pi R_{\rm m} \delta_{\rm m} \rho_0 V_{\rm ff}$ (see e.g. \cite{Elsner-Lamb-1984}, \cite{Ikhsanov-Pustilnik-1996}), in the considered case is limited to $\dot{M}_{\rm in} \sim (\delta_{\rm m}/R_{\rm m}) \dmf_{\rm c}$, where $\delta_{\rm m} = \left(D_{\rm eff}\ \tau_{\rm d}\right)^{1/2}$ is the thickness of the diffusion layer, $D_{\rm eff}$ is the effective diffusion coefficient,  $\tau_{\rm d}$ is the diffusion time and $\rho_0 = \dmf_{\rm c}/\left(4 \pi R_{\rm m}^2 V_{\rm ff}\right)$ is the density of the free-falling material at the magnetospheric boundary.
In particular, the rate of Bohm diffusion, $D_{\rm eff} \sim D_{\rm B} = \alpha_{\rm B} c k_{\rm B} T_{\rm s}/e B$, on the dynamical timescale, $\tau_{\rm d} = t_{\rm ff}$, is limited to
 \be
 \dot{M}_{\rm B} \leq 6 \times 10^{11}\,{\rm g\,s^{-1}}\ \times\ \alpha_{\rm B}^{1/2}\ \mu_{30}^{-1/14}\ m^{1/7} \dmf_{17}^{11/14}.
 \ee
which is a few orders of magnitude smaller than the mass transfer rate expected in the case of stationary accretion picture \cite{Ikhsanov-2001}.

A higher entry rate could be expected if the boundary were interchange unstable. The fastest mode is the Rayleigh-Taylor instability (see \cite{Arons-Lea-1976a}, \cite{Arons-Lea-1976b}, \cite{Elsner-Lamb-1977}, \cite{Arons-Lea-1980}). However, this instability in the considered case can occur only if the temperature of the material at the boundary is limited to $T(R_{\rm m}) \leq 0.1 - 0.3\,T_{\rm ff}(R_{\rm m})$ \cite{Arons-Lea-1976a}, which implies that cooling of the accreting material dominates heating. Otherwise, the effective acceleration applied to the material at the boundary, $g_{\rm eff} = \dfrac{G M_{\rm ns}}{R_{\rm m}^2(\theta)} \cos \chi - \dfrac{V_{\rm T_{\rm i}}^{2}}{R_{\rm curv}(\theta)}$, will be directed outwards from the star. Here $\theta$ is the magnetic latitude, $\chi$ is the angle between the radius-vector and the normal to the magnetospheric boundary, and $R_{\rm curv}$ is the curvature radius of the field lines. Studies \cite{Elsner-Lamb-1977}, \cite{Arons-Lea-1980} have shown this criterium to be satisfied only in bright X-ray pulsars with $L_{\rm x} > L_{\rm cr}$, where
 \be
L_{\rm cr} \simeq 3 \times 10^{36} \mu_{30}^{1/4} m^{1/2} R_6^{-1/8}\ {\rm erg\,s^{-1}},
 \ee
and $R_6 = R_{\rm ns}/10^6$\,cm. Otherwise, cooling time of the material at the magnetospheric boundary (which is governed by the inverse Compton scattering of hot electrons on X-ray photons emitted from the surface of the neutron star) is larger than the dynamical time on which the heating of the electrons occurs. The faint X-ray pulsars within this scenario can operate only as transient burst-like sources \cite{Lamb-etal-1977}.

The Kelvin-Helmholtz instability can be effective in fast spinning pulsars with $P_{\rm s} \sim 3\ \mu_{30}^{6/7} m^{-5/7}\ \dmf_{15}^{-3/7}\ {\rm s}$, in which the radius of magnetosphere is close to a so called co-rotation radius of the neutron star, $R_{\rm cor} = \left(GM_{\rm ns}/\omega_{\rm s}\right)^{1/3}$, at which the linear velocity at the magnetospheric boundary is equal to the Keplerian velocity \cite{Burnard-etal-1983}. Finally, the turbulent diffusion can also be effective only in fast spinning pulsars since the velocity of the turbulent motions is limited to the relative velocity between the accreting material and magnetosphere, which does not exceed $\sim \omega_{\rm s} R_{\rm m}$ \cite{Davies-Pringle-1981}.

Thus, the currently used spherical accretion scenario onto a magnetized neutron star predicts that faint long-period X-ray pulsars should be transient burst-like sources in which the accretion process onto the stellar surface occurs on the cooling time of the material at the magnetospheric boundary. This prediction contradicts observations.

The spin evolution of an accretion-powered pulsar is described by the equation
  \be\label{main}
I \dot{\omega}_{\rm s} = K_{\rm su} - K_{\rm sd},
 \ee
where $I$ is the moment of inertia of the neutron star. The spin-up torque applied to the star from the quasi-spherical accretion flow is $K_{\rm su} = \xi k_{\rm a0} \dmf_{\rm a} = \xi \Omega_{\rm orb} R_{\rm G}^2 \dmf_{\rm a}$. The spin-down torque can be evaluated by solving a model task in which the magnetosphere is approximated by a sphere of radius $R_{\rm m}$ and its interaction with surrounding material is treated in terms of rotation in viscous medium. The spin-down torque in this case is $K_{\rm sd} \sim  4 \pi R_{\rm m}^2 \nu_{\rm t} \rho_0 V_{\phi}$. Assuming the turbulent nature of the viscosity, $\nu_{\rm t} = \varepsilon_{\rm t} R_{\rm m} V_{\phi}$, where $V_{\phi} \leq \omega_{\rm s} R_{\rm m}$ is the relative azimuthal velocity between the magnetosphere and material at the boundary, one finds $K_{\rm sd} = \varepsilon_{\rm t} \mu^2/R_{\rm cor}^3$ \cite{Lipunov-1982}. Here $\varepsilon_{\rm t}$ is the efficiency parameter. Using parameters of GX\,301--2 one finds, that the spin-down torque applied to the neutron star is a factor of $1000\,\left(B/B_{\rm CRSF}\right)^2$ smaller than the spin-down torque evaluated from observations. Here $B_{\rm CRSF} = 4 \times 10^{12}$\,G is the surface field on the neutron star in this system measured through observations of CRSF. This inconsistency indicates that either the surface field of the star is a factor of 30 stronger or the accretion picture in this source differs from that expected in the currently used spherical accretion scenario.

 \section{Magnetized vs. non-magnetized accretion flow}

Spectropolarimetric observations \cite{Hubrig-etal-2006, Oksala-etal-2010, Martins-etal-2010} have shown that a relatively strong magnetization of O/B-type stars is not unusual. The strength of the large-scale field at the surface of these objects has been measured in the range $B_* \sim 100-1000$\,G, and in some cases beyond 10\,kG. The field in the environments of these objects can be evaluated by taking into account that the dipole approximation ($B \propto r^{-3}$) to the magnetic field of massive hot stars remains valid up to a distance $R_{\rm k}$, at which the dynamical pressure of their stellar wind reaches the magnetic tension. The field in the wind propagating beyond this point decreases as $B \propto r^{-2}$ \cite{Walder-etal-2011}. Therefore, the magnetic energy density in the wind at a distance $a > R_{\rm k}$ is $\msE_{\rm m}(a) = \mu_*^2/\left(2 \pi R_{\rm k}^2 a^4\right)$, which implies
  \be\label{bwa}
\msE_{\rm m}(a) \simeq 0.03\ {\rm erg\,cm^{-3}} \left(\frac{a}{10^{13}\,{\rm cm}}\right)^{-4}
 \left(\frac{\mu_*}{3 \times 10^{38}\,{\rm G\,cm^3}}\right)^2 \left(\frac{R_{\rm k}}{100\,R_{\sun}}\right)^{-2},
 \ee
where $\mu_* = (1/2) B_* R_*^3$ is the dipole magnetic moment of the massive star.

The thermal energy density of the wind at the Bondi radius, $\msE_{\rm th} = \rho_{\infty} V_{\rm s}^2$, can be evaluated by taking into account that $\rho_{\infty} = \dmf_{\rm c}/\pi R_{\rm G}^2 V_{\rm rel}$ as
 \be
\msE_{\rm th} = \frac{\dmf_{\rm c} V_{\rm rel}^3}{\pi (GM_{\rm ns})^2} V_{\rm s}^2 \simeq
0.02\ {\rm erg\,cm^{-3}} m^{-2}\ \dmf_{17}\ \left(\frac{V_{\rm rel}}{500\,{\rm km\,s^{-1}}}\right)^3
\left(\frac{V_{\rm s}}{10^6\,{\rm cm\,s^{-1}}}\right)^2.
 \ee
Here $V_{\rm s}$ is the sound speed in the wind at $R_{\rm G}$. Thus, if the neutron star is situated in a relatively slow wind the ratio $\msE_{\rm th}(R_{\rm G})/\msE_{\rm m}(R_{\rm G}) \equiv \beta$ can be about or even smaller than unity. The magnetic field of the accretion flow cannot in this case be neglected.

 \section{Magnetically controlled accretion scenario}\label{mca}

We consider spherical accretion onto a magnetized neutron star under the assumption that the parameter $\beta$ in the material captured by the star at the Bondi radius is close to unity. The ram pressure of the captured material at the Bondi radius significantly exceeds its thermal and magnetic pressure. The Alfv\'en velocity, $V_{\rm A} = B_{\rm f}/(4 \pi \rho)^{1/2}$, in the accreting material under these conditions is substantially smaller than the free-fall velocity, $V_{\rm ff}$, and, hence, the field annihilation time exceeds the dynamical time. This indicates that accretion process can initially be considered in the magnetic flux conservation approximation. Here $B_{\rm f}$ is the field strength in the accretion flow.

The magnetic field in the free-falling material is dominated by the radial component \cite{Zeldovich-Shakura-1969}, which under the condition of the magnetic flux conservation increases as $B_{\rm r}(R) \sim B_{\rm f}(R_{\rm G}) \left(r/R_{\rm G}\right)^{-2}$ \cite{Bisnovatyi-Kogan-Fridman-1970}. The magnetic pressure in the accreting material, therefore, increases as it is approaching the star as
\be
 \msE_{\rm m}(r) = \msE_{\rm m}(R_{\rm G}) \left(\frac{r}{R_{\rm G}}\right)^{-4},
 \ee
while the ram pressure of the free-falling spherical flow is $\msE_{\rm r} \propto r^{-5/2}$ (see Eq.~\ref{eram}), and thus, $\msE_{\rm m}/\msE_{\rm r} \propto r^{-3/2}$. This indicates that the gravitational energy of the neutron star in the process of spherical accretion is converted predominantly into the magnetic energy of the accreting material.

The distance $R_{\rm sh}$, at which the magnetic pressure in the accretion flow reaches the ram pressure, can be evaluated by equating $\msE_{\rm m}(R_{\rm sh}) = \msE_{\rm r}(R_{\rm sh})$. This yields \cite{Shvartsman-1971},
 \be\label{rsh}
 R_{\rm sh} = \beta^{-2/3} \left(\frac{V_{\rm s}}{V_{\rm rel}}\right)^{4/3} R_{\rm G} =
 \beta^{-2/3}\ \frac{2 GM_{\rm ns} V_{\rm s}^{4/3}}{V_{\rm rel}^{10/3}}.
 \ee
For $R_{\rm sh}$ (hereafter Shvartsman radius) to exceed Alfv\'en radius of the neutron star, the relative velocity should satisfy condition $V_{\rm rel} \leq V_{\rm mca}$, where
 \be
 V_{\rm mca} = \beta^{-1/5} \dmf_{\rm c}^{3/35} (2GM_{\rm ns})^{12/35} V_{\rm s}^{2/5} \mu^{-6/35}.
 \ee
For typical parameters of HMXBs one finds
  \bdm
V_{\rm mca} \simeq 680\ \beta^{-1/5}\ m^{12/35} \mu_{30}^{-6/35}\ \dmf_{17}^{3/35}\ V_6^{2/5}\ {\rm km\,s^{-1}},
 \edm
which under the conditions of interest substantially exceeds the value of $V_{\rm cr}$. Thus, one can distinguish a subclass of HMXBs in which the accretion occurs in spherically symmetrical fashion and the accreting material is strongly affected by the magnetic field of the flow itself. This subclass is defined by condition $V_{\rm cr} < V_{\rm rel} < V_{\rm mca}$.

Rapid amplification of the magnetic field in the spherical flow as well as deceleration of the flow by its own magnetic field at the Shvartsman radius have been confirmed by the results of numerical studies of magnetized spherical accretion onto a black hole \cite{Igumenshchev-etal-2003, Igumenshchev-2006}. These calculations have shown that the magnetized flow is shock-heated at the Shvartsman radius up to the adiabatic temperature. If the cooling of the material at $R_{\rm sh}$ is inefficient the flow switches into the convective-dominated stage in which some material is leaving the system in a form of jets and the mass accretion rate in this region proves to be smaller than its initial value.

Another important effect, the Compton cooling of the accreted material, can modify this accretion picture. Due to a large value of $V_{\rm rel}$, the Shvartsman radius in these systems is located much closer to the neutron star which is a powerful source of X-ray emission. The Compton cooling of the accreting material in this case can prevent the flow from switching into the convective-dominated stage and being ejected out from the binary system.

The plasma cooling at $R_{\rm sh}$ can be effective if the Compton cooling time \cite{Elsner-Lamb-1977},
 \be\label{tcomp}
t_{\rm c}(R_{\rm sh}) = \frac{3 \pi m_{\rm e} c^2 R_{\rm sh}^2}{2 \sigma_{\rm T} L_{\rm x}},
 \ee
is smaller than the characteristic time of plasma radial motion inside the Shvartsman radius. Here $m_{\rm e}$ is the electron mass and $\sigma_{\rm T}$ is the Thomson cross-section. If the magnetic flux in the accreting material is conserved, the accretion ends at the Shvartsman radius. Further accretion in this case is impossible. Otherwise, the magnetic energy in the flow would exceed its gravitational energy, which contradicts the energy conservation law (for discussion see \cite{Shvartsman-1971}). Therefore, the accretion flow can approach the star to a closer distance only if a dissipation of the magnetic field in the flow starts. If the field dissipation is governed by magnetic reconnection the characteristic time of the accretion process inside $R_{\rm sh}$ is limited to $t \geq t_{\rm rec}$, where
 \be\label{trec}
 t_{\rm rec} = \frac{r}{\eta_{\rm m} V_{\rm A}} = \eta_{\rm m}^{-1} t_{\rm ff}
 \left(\frac{V_{\rm ff}}{V_{\rm A}}\right).
 \ee
Theoretical studies \cite{Parker-1971} and observations \cite{Noglik-etal-2005} suggest the value of the efficiency parameter $\eta_{\rm m}$ to be in the range of 0.01--0.15. It should be also mentioned that the Alfv\'en velocity in the accretion flow reaches the free-fall velocity at the Shvartsman radius. Combining Eqs.~(\ref{tcomp}) and (\ref{trec}), one can express inequality $t_{\rm c} \leq t_{\rm rec}$ as
 \be
 L_{\rm x} \geq \eta_{\rm m} L_{\rm cr} \left(\frac{R_{\rm sh}}{R_{\rm A}}\right)^{1/2}.
 \ee
This indicates that the cooling of magnetized flow can be effective even in faint X-ray pulsars in which the Shvartsman radius does not significantly exceed the Alfv\'en radius of the neutron star.

The accretion picture of a cold magnetized gas has been discussed in \cite{Bisnovatyi-Kogan-Ruzmaikin-1974, Bisnovatyi-Kogan-Ruzmaikin-1976}. It has been shown that the material in this case tends to flow along the field lines and in the region $r \leq R_{\rm sh}$ accumulates in a dense non-Keplerian slab (see Fig.\,1 in \cite{Bisnovatyi-Kogan-Ruzmaikin-1976}). The material in the slab is confined by the magnetic field of the flow itself and its radial motion continues as the field is annihilating. The accretion process in the slab, therefore,  occurs on the timescale of $t_{\rm rec}$. This indicates that the thermal pressure in the slab is equal to magnetic pressure and hence, the gas density at the inner radius of the slab, which is equal to the magnetospheric radius, can be evaluated as
  \be\label{rhosl}
  \rho_{\rm sl} = \frac{\mu^2 m_{\rm p}}{2 \pi k_{\rm B} T_0 R_{\rm m}^6}.
  \ee
Here $T_0$ is the gas temperature at the inner radius of the slab. The thickness of the slab depends on the magnetic field configuration and in the first approximation can be evaluated from the continuity equation as
 \be\label{hz}
 h_{\rm z}(R_{\rm m}) = \frac{\dmf }{4 \pi \eta_{\rm m} R_{\rm m} \rho_{\rm sl} V_{\rm ff}}.
 \ee

The rate of plasma diffusion from the slab into the stellar field can be evaluated as
 \be\label{dmfin-1}
   \dmf_{\rm in}(R_{\rm m}) = 4 \pi R_{\rm m} \delta_{\rm m} \rho_{\rm sl} V_{\rm ff} = 4 \pi R_{\rm m}^{3/2} D_{\rm eff}^{1/2} \rho_{\rm sl} V_{\rm ff}^{1/2},
  \ee
where $\delta_{\rm m} = \left(D_{\rm eff}\ \tau_{\rm d}\right)^{1/2}$ and the diffusion time, $\tau_{\rm d} = t_{\rm ff}(R_{\rm m}) = R_{\rm m}/V_{\rm ff}(R_{\rm m})$, is equal to the dynamical time on which the material inside the magnetosphere leaves the diffusion layer by flowing with the free-fall velocity along the stellar field lines towards the stellar surface. As long as $\dmf_{\rm in}$ is smaller than the mass capture rate and, correspondingly, the rate of mass transfer in the slab in the radial direction, $\dmf_{\rm a}$, material accumulates at the inner radius of the slab and the gas pressure in this region increases. The inner radius of the slab is defined by equating the gas pressure to the magnetic pressure by the dipole field of the neutron star. Therefore, as the material at the slab accumulates, its inner radius decreases and the material is getting closer to the star. A stationary accretion picture is expected as the magnetospheric radius reaches a value at which $\dmf_{\rm in}(R_{\rm m}) = \dmf_{\rm a}$. This value depends on the nature of the diffusion process and, therefore, on the value of the effective diffusion coefficient, which in the considered case can be limited to $D_{\rm B} \leq D_{\rm eff} \leq D_{\rm t}$, where $D_{\rm B}$ and $D_{\rm t}$ are the Bohm and turbulent diffusion coefficients, respectively.
\begin{description}
\item[Bohm diffusion]
  The Bohm diffusion coefficient is $D_{\rm B} = \alpha_{\rm B} \left(c k_{\rm B} T_0 R_{\rm m}^3/2 e \mu \right)$, where $\alpha_{\rm B}$ is the efficiency parameter, which according to \cite{Gosling-etal-1991} lies in the range $0.1-0.25$. Putting this value to Eq.~(\ref{dmfin-1}), combining it with Eq.~(\ref{rhosl}) and solving equation $\dmf_{\rm in}(R_{\rm m}) = \dmf_{\rm a}$ for $R_{\rm m}$, one finds
  \be\label{rmb}
  R_{\rm m}^{\rm (B)} \simeq 7 \times 10^7\ {\rm cm}\ \times \ \alpha_{0.1}^{2/13}\ \mu_{30}^{6/13}\ T_6^{-2/13}\ m^{5/13}\ L_{37}^{-4/13}\ \left(\frac{R_{\rm ns}}{10\,{\rm km}}\right)^{-4/13},
  \ee
where $\alpha_{0.1}=\alpha/0.1$. The plasma temperature at the magnetospheric boundary, $T_6$, is normalized here according to \cite{Masetti-etal-2006} and expressed in units $10^6$\,K. Finally, $L_{37}$ is the X-ray luminosity of the pulsar in units $10^{37}\,{\rm erg\,s^{-1}}$.
\item[Turbulent diffusion]
 The turbulent diffusion coefficient can be evaluated as $D_{\rm t} = \alpha_{\rm t} V_{\rm s} h_{\rm z}$, where $\alpha_{\rm t}$ is the efficiency parameter of the turbulence in the slab. Putting this value to Eq.~(\ref{dmfin-1}), combining it with Eq.~(\ref{hz}) and solving equation $\dmf_{\rm in}(R_{\rm m}) = \dmf_{\rm a}$ for $R_{\rm m}$, one finds
  \be
  R_{\rm m}^{\rm (t)} \simeq 10^9\ {\rm cm}\ \times\ \alpha_{\rm t}^{1/4} \eta_{\rm m}^{-1/4} \mu_{30}^{1/2} m^{1/4} T_6^{-1/8} L_{37}^{-1/4} \left(\frac{R_{\rm ns}}{10\,{\rm km}}\right)^{-1/4}.
  \ee
\end{description}
The value of $R_{\rm m}^{\rm (B)}$ is smaller than the Alfv\'en radius of the neutron star evaluated in the non-magnetized accretion flow scenario, $R_{\rm A}$, by a factor of 5, but remains substantially larger than the radius of the neutron star, $R_{\rm ns} \simeq 10$\,km for any reasonable value of the X-ray luminosity. The condition $R_{\rm m}^{\rm (t)} \leq R_{\rm A}$ is satisfied if $\left(\alpha_{\rm t}/\eta_{\rm m}\right)\,L_{37}^{-1/28} \leq 0.02$. This indicates than the plasma entry into the stellar field in the Magnetically Controlled Accretion (MCA) scenario can be explained in terms of diffusion and no additional assumptions about instability of the magnetospheric boundary are required.

 \section{Torque applied to neutron stars in MCA scenario}

For an accreting slowly-rotating neutron star to spin down, the angular velocity of the accreting material at the magnetospheric boundary should be smaller that the angular velocity of the star itself, i.e. $\omega_{\rm en} < \omega_{\rm s}$ \cite{Bisnovatyi-Kogan-1991}. This condition in the case of GX\,301--2 is satisfied only if
 \bdm
\xi < 0.001 \left(\frac{P_{\rm s}}{685\,{\rm s}}\right)^{-1} \left(\frac{P_{\rm orb}}{41.5\,{\rm d}}\right)  \left(\frac{R_{\rm G}}{2.5 \times 10^{11}\,{\rm cm}}\right)^2 \left(\frac{R_{\rm m}^{\rm (B)}}{10^8\,{\rm cm}}\right)^{-2}
 \edm
This is two orders of magnitude smaller than the average value of this parameter derived in numerical studies of spherical accretion in non-magnetized flow approximation \cite{Ruffert-1999}. On the other hand, a significant dissipation of angular momentum is expected in the case of a magnetized flow (see e.g. \cite{Mestel-1959}). This finding, therefore, favors the MCA scenario in this pulsar.

The absolute value of spin-up torque applied to the neutron star from the accreting material, $K_{\rm su} = \xi \Omega_{\rm orb} R_{\rm G}^2 \dmf_{\rm a}$, in this case, is a factor of 60 smaller than the absolute value of the spin-down torque inferred from observations \cite{Doroshenko-etal-2010},
 \be\label{k1}
K_{\rm sd} = 2 \pi I |\dot{\nu}_{\rm sd}| \simeq 6 \times 10^{32}\ I_{45} \left(\frac{\dot{\nu}_{\rm
sd}}{10^{-13}\,{\rm Hz\,s^{-1}}}\right) {\rm dyne\,cm},
 \ee
and hence, it does not affect the spin behavior of the pulsar during this stage. Here $I_{45}$ is the moment of inertia of the neutron star in units of $10^{45}\,{\rm g\,cm^2}$.

The spin-down of an accreting neutron star is associated with interaction between the magnetosphere co-rotating with the star and the material at the magnetospheric boundary. Since both the Reynolds number and magnetic Reynolds number in the slab are very large, this interaction leads to turbulization of the material at the boundary. The spin-down torque associated with this process can be evaluated as $K_{\rm sd}^{(t)} = \varepsilon_{\rm t} \dmf_{\rm eff} \omega_{\rm s} R_{\rm m}^2$, where $\dmf_{\rm eff}$ is the mass involved into the turbulent motions in a time unit, and $\varepsilon_{\rm t}$ is the efficiency parameter. The energy transferred by the turbulent motions in a time unit is, correspondingly, $\dot{E}_{\rm t} = \omega_{\rm s} K_{\rm sd}^{(t)} = \varepsilon_{\rm t} \dmf_{\rm eff} V_{\phi}^2(R_{\rm m})$.

Parameter $\dmf_{\rm eff}$ can be evaluated by equating the ram pressure of the material in the slab with the magnetic pressure due to the dipole field of the neutron star, that is
 \bdm
\dmf_{\rm eff} = \frac{\mu^2}{R_{\rm m}^{7/2} (2 GM_{\rm ns})^{1/2}}.
 \edm
Thus, the spin-down torque applied to the neutron star surrounded by the slab is
 \be\label{ksdt}
 K_{\rm sd}^{(t)} = \frac{\varepsilon_{\rm t}\ \mu^2\ \omega_{\rm s}}{R_{\rm m}^{3/2} (2 GM_{\rm ns})^{1/2}}.
 \ee
Putting $R_{\rm m} = R_{\rm m}^{(B)}$ into this equation and using parameters of GX\,301--2 one finds
 \bdm
 K_{\rm sd}^{(t)} \simeq 2 \times 10^{33} \varepsilon_{\rm t} \alpha_{0.1}^{-3/13} P_{685}^{-1} T_6^{3/13} L_{37}^{6/13} m^{-14/13} \left(\frac{B}{B_{\rm CRSF}}\right)^{17/13} \left(\frac{R_{\rm ns}}{10\,{\rm km}}\right)^{57/13}\ {\rm dyne\,cm},
 \edm
where $P_{685}$ is the spin period of the pulsar in the units of 685\,s, and $B_{\rm CRSF} = 4 \times 10^{12}$\,G is the strength of the surface field of the neutron star measured through observations of the Compton Resonant Scattering Feature in its X-ray spectrum ( see \cite{La-Barbera-etal-2005}, and references therein). Thus, the spin-down torque applied to the neutron star in GX\,301--2 during the spin-down trends of the pulsar can be explained in terms of the MCA scenario provided the turbulent efficiency is $\varepsilon_{\rm t} \geq 0.3$, which  is in a good agreement with theoretically predicted value (see e.g. \cite{Landau-Lifshits-2000}).

 \section{Conclusions}

The currently used scenario of spherical accretion in HMXBs is solely built around the assumption that the accretion flow is non-magnetized. We show this assumption to be controversial. Recent results of surface field measurement of O/B-type stars suggest that the magnetic field energy density in the material captured by neutron stars in wind-fed HMXBs is comparable with (or even higher than) the thermal energy density of the wind at the Bondi radius. The accretion flow in this case can be strongly affected by its own magnetic field. The magnetic field in the spherical accretion flow is rapidly growing. The flow is decelerated as the magnetic pressure due to magnetic field of the flow itself reaches its ram pressure. The distance to the deceleration region (Shvartsman radius) under certain conditions exceeds the Alfv\'en radius of the neutron star. The characteristic time of accretion process inside the Shvartsman radius is limited to the time of annihilation of the magnetic field in the accretion flow, which under the conditions of interest substantially exceeds the dynamical time as well as the cooling time of the accretion flow in this region. The accreting material in this case accumulates around the magnetosphere of the neutron star forming a dense non-Keplerian slab. The entry of the material from the slab into the magnetosphere is governed by turbulent diffusion. The pulsar spin evolution in this Magnetically Controlled Accretion (MCA) scenario is determined by the process of angular momentum exchange between the star and the slab. The rate of angular momentum transfer evaluated in this scenario is in good agreement with observations.


\begin{theacknowledgments}
We would like to thank G.S.\,Bisnovatyi-Kogan, M.H.\,Finger and M.V.\,Medvedev for useful discussions. NRI and LAP thank the organizing committee for kind hospitality and French Embassy in Israel (Office of Science and Technology) and Ben-Gurion University of the Negev for support in attending the Workshop. The research has been partly supported by the Program of RAS Presidium N\,19, NSH-3645.2010.2, and by the grant ``Infrastructure'' of Israel Ministry of Science.
\end{theacknowledgments}


\begin{thebibliography}{99}

\bibitem[Arons and Lea(1976a)]{Arons-Lea-1976a}
 Arons J., Lea S.M., \emph{Astrophys. J.}, \textbf{207}, 914--936 (1976).

\bibitem[Arons and Lea(1976b)]{Arons-Lea-1976b}
 Arons J., Lea S.M., \emph{Astrophys. J.}, \textbf{210}, 792--804 (1976).

\bibitem[Arons and Lea(1980)]{Arons-Lea-1980}
 Arons J., Lea S.M., \emph{Astrophys. J.}, \textbf{235}, 1016--1037 (1980).

\bibitem[Bisnovatyi-Kogan and Fridman(1970)]{Bisnovatyi-Kogan-Fridman-1970}
 Bisnovatyi-Kogan, G.S., Fridman, A.M., \emph{Soviet Astronomy}, \textbf{13}, 566--568 (1970).

\bibitem[Bisnovatyi-Kogan and Ruzmaikin(1974)]{Bisnovatyi-Kogan-Ruzmaikin-1974}
 Bisnovatyi-Kogan, G.S., Ruzmaikin, A.A., \emph{Astrophys. and Space Sci.}, \textbf{28}, 45--59 (1974).

\bibitem[Bisnovatyi-Kogan and Ruzmaikin(1976)]{Bisnovatyi-Kogan-Ruzmaikin-1976}
  Bisnovatyi-Kogan, G.S., Ruzmaikin, A.A., \emph{Astrophys. and Space Sci.}, \textbf{42}, 401--424 (1976).

\bibitem[Bisnovatyi-Kogan(1991)]{Bisnovatyi-Kogan-1991}
  Bisnovatyi-Kogan, G.S., \emph{Astron. Astrophys.}, \textbf{245}, 528--530 (1991).

\bibitem[Burnard et~al.(1983)]{Burnard-etal-1983}
 Burnard, D.J., Lea, S.M., Arons, J., \emph{Astrophys. J.}, \textbf{266}, 175--187 (1983).

\bibitem[Davies and Pringle(1981)]{Davies-Pringle-1981}
 Davies, R.E., Pringle, J.E., \emph{Mon. Not. R. Astron. Soc.}, \textbf{196}, 209--224 (1981).

\bibitem[Doroshenko et~al.(2010)]{Doroshenko-etal-2010}
 Doroshenko, V., Santangelo, A., Suleimanov, V., et~al., \emph{Astron. Astrophys.}, \textbf{515}, 10--20 (2010).

\bibitem[Elsner and Lamb(1977)]{Elsner-Lamb-1977}
 Elsner R.F., Lamb F.K., \emph{Astrophys. J.}, \textbf{215}, 897--913 (1977).

\bibitem[Elsner and Lamb(1984)]{Elsner-Lamb-1984}
 Elsner R.F., Lamb F.K., \emph{Astrophys. J.}, \textbf{278}, 326--344 (1984).

\bibitem[Gosling et~al.(1991)]{Gosling-etal-1991}
  Gosling J.T., Thomsen M.F., Bame S.J., et~al., 1991, \emph{J. Geophys. Res.}, \textbf{96}, 14097--14106
  (1991)

\bibitem[Hubrig et~al.(2006)]{Hubrig-etal-2006}
 Hubrig, S., Yudin, R.V., Sch\"oller, M., Pogodin, M.A., \emph{Astron. Astrophys.}, \textbf{446}, 1089--1094  (2006).

\bibitem[Igumenshchev et~al.(2003)]{Igumenshchev-etal-2003}
 Igumenshchev, I.V., Narayan, R., Abramowicz, M.A., \emph{Astrophys. J.}, \textbf{592}, 1042--1059 (2003).

\bibitem[Igumenshchev(2006)]{Igumenshchev-2006}
 Igumenshchev, I.V., \emph{Astrophys. J.}, \textbf{649}, 361--372 (2006).

\bibitem[Ikhsanov(2001)]{Ikhsanov-2001}
 Ikhsanov, N.R., \emph{Astron. Astrophys.}, \textbf{375}, 944--949 (2001)

\bibitem[Ikhsanov(2007)]{Ikhsanov-2007}
 Ikhsanov, N.R., \emph{Mon. Not. R. Astron. Soc.}, \textbf{375}, 698--704 (2007)

\bibitem[Ikhsanov and Pustilnik(1996)]{Ikhsanov-Pustilnik-1996}
 Ikhsanov, N.R., Pustil'nik, L.A., \emph{Astron. Astrophys.}, \textbf{312}, 338--344 (1996).

\bibitem[Koh et~al.(1997)]{Koh-etal-1997}
 Koh, D.T., Bildsten, L., Chakhrabarty, D., et~al., \emph{Astrophys. J.}, \textbf{479}, 933--947 (1997).

\bibitem[La-Barbera et~al.(2005)]{La-Barbera-etal-2005}
 La Barbera, A., Segreto, A., Santangelo, A., et~al., \emph{Astron. Astrophys.}, \textbf{438}, 617--632 (2005).

\bibitem[Lamb et~al.(1977)]{Lamb-etal-1977}
 Lamb, F.K., Fabian, A.C., Pringle, J.E., Lamb, D.Q., \emph{Astrophys. J.}, \textbf{217}, 197--212 (1977).

\bibitem[Landau and Lifshits(2000)]{Landau-Lifshits-2000}
 L.D. Landau, E.M. Lifshitz, \emph{Fluid Mechanics}, Butterworth-Heinemann, Oxford, 2000, pp. 95--153

\bibitem[Liu etal.(2006)]{Liu-etal-2006}
 Liu, Q.Z., van Paradijs, J., van den Heuvel, E.P.J., \emph{Astron. Astrophys.}, \textbf{455}, 1165--1168 (2006).

\bibitem[Lipunov(1982)]{Lipunov-1982}
 Lipunov, V.M., \emph{Soviet Astronomy}, \textbf{26}, 537--541 (1982)

\bibitem[Martins et~al.(2010)]{Martins-etal-2010}
  Martins, F., Donati, J.-F., Marcolino, W.L.F., et~al., \emph{Mon. Not. R. Astron. Soc.}, \textbf{407}, 1423--1432 (2010).

\bibitem[Masetti et~al.(2006)]{Masetti-etal-2006}
 Masetti, N., Orlandini, M., Dal Fiume, D., et~al., \emph{Astron. Astrophys.}, \textbf{445}, 653--660 (2006).

\bibitem[Mestel(1959)]{Mestel-1959}
 Mestel, L., \emph{Mon. Not. R. Astron. Soc.}, \textbf{119}, 223--248 (1959).

\bibitem[Michel(1977)]{Michel-1977}
 Michel F.C., \emph{Astrophys. J.}, \textbf{213}, 836--839 (1977).

\bibitem[Noglik et~al.(2005)]{Noglik-etal-2005}
 Noglik, J.B., Walsh, R.W., Ireland, J., \emph{Astron. Astrophys.}, \textbf{441}, 353--360 (2005).

\bibitem[Oksala et~al.(2010)]{Oksala-etal-2010}
 Oksala, M.E., Wade, G.A., Marcolino, et~al., \emph{Mon. Not. R. Astron. Soc.}, \textbf{405}, L51--L55 (2010).

\bibitem[Parker(1971)]{Parker-1971}
 Parker, E.N., \emph{Astrophys. J.}, \textbf{163}, 279--285 (1971).

\bibitem[Ruffert(1999)]{Ruffert-1999}
 Ruffert, M., \emph{Astron. Astrophys.}, \textbf{346}, 861--877 (1999).

\bibitem[Shvartsman(1971)]{Shvartsman-1971}
 Shvartsman, V.F., \emph{Soviet Astronomy}, \textbf{15}, 377--384 (1971).

\bibitem[Walder et~al.(2011)]{Walder-etal-2011}
Walder, R., Folini, D., Meynet, G., \emph{Space Sci. Rev.}, \textbf{125}, in press (2011).

\bibitem[Zeldovich and Shakura(1969)]{Zeldovich-Shakura-1969}
 Zel'dovich, Ya.B., Shakura, N.I., \emph{Soviet Astronomy}, \textbf{13}, 175--183 (1969).
\end{thebibliography}
\end{document}